\documentclass[aps,pra,reprint,superscriptaddress, nofootinbib]{revtex4-1}
\usepackage{amsmath, amsthm, amssymb, amsfonts}
\usepackage{graphicx, color}
\usepackage{braket}
\usepackage{xcolor}
\usepackage{hyperref}
\newcommand{\ot}{\otimes}
\newcommand{\op}{\oplus}
\newcommand{\wt}{\widetilde}
\newcommand{\ovl}{\overline}
\newcommand{\Tr}{\mathrm{Tr}}
\newcommand{\CH}{\mathcal H}
\newcommand{\Hc}{\CH_{code}}
\newcommand{\Hp}{\CH_{phys}}
\newcommand{\al}{\alpha}
\newcommand{\be}{\beta}
\newcommand{\Ao}{A^\al_1}
\newcommand{\Aob}{\ovl A^\al_1}
\newcommand{\At}{A^{\al\be}_2}
\newcommand{\Atb}{\ovl A^{\al\be}_2}
\newtheorem{thrm}{Theorem}

\begin{document}

\title{\bf The Ryu-Takayanagi Formula from Quantum Error Correction: An Algebraic Treatment of the Boundary CFT}
\author{Helia Kamal}
\affiliation{Department of Physics, University of California, Berkeley, CA 94720, USA}
\author{Geoffrey Penington}
\affiliation{Stanford Institute for Theoretical Physics, Stanford University, CA 94305, USA}

\begin{abstract}
It was recently shown by Harlow that any quantum error correcting code, satisfying the same complementary recovery properties as AdS/CFT, will obey a version of the Ryu-Takayanagi formula. In his most general result, Harlow allowed the bulk algebras to have nontrivial center, which was necessary for the ``area operator'' in this Ryu-Takayanagi formula to be nontrivial. However, the boundary Hilbert space was still assumed to factorise into Hilbert spaces associated with complementary boundary regions. We extend this work to include more general boundary theories, such as gauge theories, where the subalgebras associated with boundary regions may also have nontrivial center. We show the equivalence of a set of four conditions for a bulk algebra to be reconstructable from a boundary algebra, and then show that complementary recovery implies that the algebraic boundary entropy obeys a Ryu-Takayanagi formula. In contrast, we show that the distillable boundary entropy does not obey any such formula. If an additional ``log dim R'' term is added to the algebraic entropy, it will still obey a Ryu-Takayanagi formula, with a different area operator. However, since the ``log dim R'' term is a sum over local boundary contributions, we argue that it can only be related to the regularisation of the area at the bulk cut-off.
\end{abstract}

\maketitle

\section{Introduction}


By establishing a duality between certain conformal field theories and quantum gravity in asymptotically anti-de Sitter geometries, the AdS/CFT correspondence \cite{maldacena1999large,witten1998anti} allows us to learn about theories of quantum gravity by studying their better understood CFT duals. However, more than a decade after this correspondence was discovered, many aspects of it remained mysterious. If AdS/CFT was an isomorphism between two dual theories, how could a single bulk operator correspond to multiple boundary operators? How did an operator in the bulk commuting with all local operators on the boundary not violate the time-slice axiom? How could boundary entropy, a nonlinear quantity, be proportional to bulk area, a linear observable, as claimed by the Ryu-Takayanagi formula~\cite{ryu2006holographic,ryu2006aspects}?

In 2014, Almheiri, Dong, and Harlow resolved many of these puzzles by reinterpreting the bulk-to-boundary map as an example of a quantum error correcting code \cite{almheiri2015bulk}. In this interpretation, a smaller bulk Hilbert space, of states with `nice' semiclassical bulk description, is embedded as a `code subspace' in a larger boundary CFT Hilbert space. Bulk operators commute with all local boundary operators because we can error correct the erasure of any sufficiently small boundary region; the multiple boundary operators that correspond to a single bulk operator are the multiple physical operators that can correspond to a single logical operator. This discovery provided a new connection between quantum information theory and quantum gravity and led to many other significant developments at the intersection of these fields~\cite{mintun2015bulk,pastawski2015holographic,hayden2016holographic, verlinde2017emergent, hayden2018learning, penington2019entanglement, almheiri2019entropy}.

In particular, later work by Harlow \cite{harlow2017ryu} made it clear that the Ryu-Takayanagi formula, including the FLM correction \cite{faulkner2013quantum}, is not just consistent with the quantum error correction story, but is, in fact, a feature of \emph{any} quantum error correcting code with the correct properties. 

Harlow studied a series of increasingly sophisticated models of the bulk to boundary map. In the most sophisticated model, bulk regions are not associated with tensor product factors of the bulk Hilbert space. Instead, they are associated with finite-dimensional von Neumann subalgebras of the bulk Hilbert space. The error correction properties of AdS/CFT, specifically entanglement wedge reconstruction \cite{czech2012gravity, wall2014maximin, headrick2014causality, dong2016reconstruction, jafferis2016relative, faulkner2017bulk, cotler2019entanglement}, then imply a version of the Ryu-Takayanagi formula, where the `area operator' is some non-trivial linear operator in the \emph{center} of the bulk algebra associated with the entanglement wedge.

However, even in this most sophisticated model, Harlow still assumed that the boundary CFT Hilbert space could be decomposed as the tensor product of smaller Hilbert spaces associated with each boundary region. Formally, this assumption is untrue for any continuum CFT \cite{witten2018notes}. Heuristically, this is because any `product state', with no entanglement between two complementary boundary regions, would have infinite energy and so is not in the CFT Hilbert space. Instead, boundary regions are associated with infinite-dimensional von Neumann subalgebras.

For sufficiently simple conformal field theories, such as free theories, the von Neumann subalgebras will be factors, i.e. have trivial center. Such algebras are morally equivalent to the algebra of operators acting on a subsystem Hilbert space in finite dimensions. If we regulate our CFT in the UV, as is necessary to make each side of the Ryu-Takayanagi formula finite, the boundary Hilbert space will indeed factorise as a product of subregion Hilbert spaces.

However, this is not true for the most prominent examples of the AdS/CFT correspondence, such as $N=4$ super-Yang Mills theory. In such cases, the boundary theory has a gauge symmetry, which causes the physical Hilbert space of gauge-invariant states to never factorise. Instead, the subalgebras associated to boundary regions have a nontrivial center, just like the bulk algebras, even when we regulate the theory in the UV.

In this paper, we extend the arguments of Harlow \cite{harlow2017ryu} to account for this fact. We first show the equivalence of a set of four conditions for the reconstructability of a bulk subalgebra from a boundary subalgebra.

We then show that, whenever a bulk algebra $M$ can be reconstructed from a boundary algebra $N$ and the commutant $M'$ of the bulk algebra can be reconstructed from the commutant $N'$ of the boundary algebra, the \emph{algebraic} entropy of the boundary algebra $N$ satisfies a Ryu-Takayanagi formula
\begin{align}
S(\wt \rho, N) &= \Tr (\wt \rho {\cal L}) + S(\wt \rho, M)  
\end{align}
where $S(\wt \rho, N)$ is the algebraic entropy \cite{casini2014remarks} of the boundary algebra $N$, $S(\wt \rho, M)$ is the algebraic entropy of the bulk algebra $M$ and $\cal L$ is a linear operator that corresponds to area in holographic theories.\footnote{In related work, it was shown in \cite{kang2018holographic} that algebraic reconstruction is equivalent to an equality between bulk and boundary relative entropies, even for infinite-dimensional von Neumann algebras. However, it is hard to define a Ryu-Takayanagi formula for infinite-dimensional algebras, since the algebraic entropy is divergent. We also note that condition (i) of Theorem \ref{thrm:conditions} is not expected to generalize to the infinite-dimensional setting.}

In contrast, an alternative definition of the entropy of an algebra, the distillable entropy \cite{ghosh2015entanglement, soni2016aspects}, which agrees with the algebraic entropy for the algebra of operators on a subsystem (as considered in \cite{harlow2017ryu}), does \emph{not} satisfy a Ryu-Takayanagi formula. 

A third definition of entanglement entropy, defined specifically for non-Abelian gauge theories such as $N=4$ super-Yang Mills, is the extended Hilbert space, or ``$\log \text{dim}\, R$'', entropy considered in \cite{donnelly2012decomposition, donnelly2014entanglement, lin2018comments}. This entropy differs from the algebraic entropy only by a linear term, and so also satisfies a Ryu-Takayanagi formula. Since the additional linear term is a sum over local boundary observables, we argue that this additional term is only relevant to the regulation of the bulk area at the cut-off scale, and that both the algebraic and ``$\log \text{dim}\, R$'' entropies obey the ``real'' Ryu-Takayanagi formula.

\textbf{Notation}. The notation for this work is inevitably heavy at times. Therefore, to make it easier for the reader to follow the series of generalizations, we borrow the notation of \cite{harlow2017ryu} with minor modifications. We will use upper case Roman letters to label the subsystems of the physical Hilbert space $\Hp$, such as $\CH_{A_\be}, \CH_{\ovl A_\be}$, etc, and lower case letters for the subsystems of the code subspace $\Hc$, such as $\CH_{a_\al}, \CH_{\ovl a_\al}$, etc. (Greek letters are used to refer to superselection sectors). We will use the ``tilde" symbol on states and operators in $\Hc$, e.g. $\ket{\wt \psi}$ and $\wt O$. When working with von Neumann algebras, the ``prime" symbol will be used to distinguish operators in the commutant from those in the algebra itself. Moreover, we will use the ``overline" symbol to distinguish the subsystems that the commutant acts on from those acted on by the algebra.

\section{Preliminaries} \label{prelim}
In this section, we briefly review some properties of the AdS/CFT correspondence and basic definitions of von Neumann algebras that we use throughout our work.

\subsection{AdS/CFT background}
The AdS/CFT correspondence is the statement that certain conformal field theories in $d$ dimensions are dual to some quantum theory of gravity in $(d+1)$-dimensional asymptotically anti-de Sitter space-time. The duality is commonly visualized on a cylinder, with an AdS geometry in the interior of the cylinder described by the `bulk' quantum gravity theory, and the dual `boundary' CFT living on the cylinder itself. 

One of the fascinating features of the duality is the connection between bulk geometry and boundary entanglement. This relationship was quantified by Ryu and Takayanagi \cite{ryu2006holographic,ryu2006aspects} and allows for the calculation of entanglement entropies in the boundary CFT at leading order using simple geometrical calculations in the bulk. In the semiclassical limit where Newton's constant $G_N$ is small, the entanglement entropy of any subregion $A$ of the boundary, is given, up to small $O(G_N)$ corrections, by the Ryu-Takayanagi (RT) formula as $$S({A}) = \frac{\text{Area}(\gamma_A)}{4G_N} + S_\text{bulk},$$ where $\gamma_A$, called the \emph{RT surface}, is the surface of minimal area in the bulk that is homologous to $A$.\footnote{This statement is valid for static spacetimes. For more general spacetimes one needs to use the HRT prescription \cite{hubeny2007covariant, wall2014maximin}. When quantum corrections are important, one should use the quantum extremal surface prescription \cite{engelhardt2015quantum, dong2018entropy}.} The RT surface divides the bulk into two regions: one that is enclosed by $A$ and $\gamma_A$, and one that is not. The former is called the \emph{entanglement wedge} of $A$ and is denoted by ${\cal E}_A$.\footnote{More precisely, the entanglement wedge is the bulk domain of dependence of this region.} $S_\text{bulk}$ refers to the bulk entanglement entropy between the two regions. We summarize the above in the following precise statement of the RT formula using the language of states and operators.

\textbf{Ryu-Takayanagi Formula}: Given a state $\rho$ in the CFT and a boundary subregion $A$, the von Neumann entropy of the reduced state $\rho_A$ obeys 
\begin{align} \label{RT}
S(\rho_A) = \Tr (\rho {\cal L}_A) + S(\rho_{{\cal E}_A}),
\end{align}
where the operator ${\cal L}_A \equiv \frac{Area(\gamma_A)}{4G_N} + ...$ calculates the area of the RT surface up to $O(G_N)$ corrections.

Another interesting feature of the duality is the mapping between bulk and boundary operators. It has long been known that bulk operators can have multiple different boundary representations. In fact, any bulk operator has representations that act only on part of the boundary. The question of which part of the bulk can be reconstructed on a given boundary subregion has been studied in detail in \cite{czech2012gravity, wall2014maximin, headrick2014causality, dong2016reconstruction, jafferis2016relative, faulkner2017bulk, cotler2019entanglement, chen2019entanglement} and can be summarized in the following statement of entanglement wedge reconstruction:

\textbf{Entanglement Wedge Reconstruction}: Given a subregion $A$ of the boundary, a bulk operator can be reconstructed on $A$ if and only if it lies in the entanglement wedge ${\cal E}_A$.

\subsection{von Neumann algebras on finite-dimensional Hilbert spaces} \label{VNA}
In this section, we review the basic terminology of von Neumann algebras on finite dimensional Hilbert spaces. The reader is encouraged to refer to the appendix of \cite{harlow2017ryu} for a more comprehensive introduction.

A \textit{von Neumann algebra} $M$, acting on a Hilbert space $\CH$ is a subset of the linear operators acting on $\CH$ that contains all scalar multiples of the identity and is closed under addition, multiplication, and Hermitian conjugation. Any von Neumann algebra $M$ on $\CH$ naturally induces two other von Neumann algebras on $\CH$: the \textit{commutant}, denoted $M'$, is the set of linear operators in $\CH$ that commute with all operators in $M$, while the \textit{center}, denoted $M_C$, is the intersection of $M$ and its commutant $M'$, i.e. the set of all linear operators in $M$ that commute with every single operator in $M$.

Every von Neumann algebra $M$ on $\CH$ induces a unique Hilbert space decomposition of the form $\CH = \op_\al (H_{A_\al} \ot H_{\ovl A_\al})$ such that the operators in $M$, $M'$, and $M_C$ take the following form:
\begin{align}
O \in M &\rightarrow O = \op_\al (O_{A_\al} \ot I_{\ovl A_\al}) \nonumber \\
O' \in M' &\rightarrow O' = \op_\al (I_{A_\al} \ot O'_{\ovl A_\al})  \\
O_C \in M_C &\rightarrow O_C = \op_\al \lambda_\al (I_{A_\al} \ot I_{\ovl A_\al}) \nonumber
\end{align}
That is, all operators in $M$, $M'$, and $M_C$ are block-diagonal in $\al$, and within each block, $M$ only acts nontrivially on $A_\al$, $M'$ only acts nontrivially on $\ovl A_\al$, and $M_C$ is proportional to identity. In the special case where $|\al| = 1$, i.e. $M_C$ is trivial, $M$ is called a \textit{factor}.

Given a state $\rho$ and a von Neumann algebra $M$ on $\CH$, there exists a state $\rho_M \in M$ such that ${\mathbb E}_\rho (O)={\mathbb E}_{\rho_M}(O)$ for all $O \in M$, where ${\mathbb E}_\rho (O) = \Tr (\rho O)$. To find $\rho_M$, we first note that any state $\rho \in \CH$ can be written in block form with respect to the direct sum decomposition $\CH = \op_\al (H_{A_\al} \ot H_{\ovl A_\al})$ induced by $M$. We also note that since all operators in $M$ are block diagonal, only the diagonal blocks of $\rho$ will contribute to the expectation values of the operators and we can ignore the rest. We will then have
\begin{align}
\rho =
\begin{pmatrix}
p_1 \rho_{A_1 \ovl A_1} && \cdots && \cdots\\
\vdots && p_2 \rho_{A_2 \ovl A_2} &&\cdots \\
\vdots && \vdots && \ddots
\end{pmatrix},
\end{align}
where we have chosen $p_\al \in [0,1]$ so that $\Tr(\rho_{A_\al \ovl A_\al})=1$. We can now define
\begin{align} \label{rhoM}
\rho_M = \op_\al \left(p_\al \rho_{A_\al} \ot \frac{I_{\ovl A_\al}}{|\ovl A_\al|}\right),
\end{align}
where $\rho_{A_\al} = \Tr_{\ovl A_\al} (\rho_{A_\al \ovl A_\al})$. Given this definition of state, the entropy of the state $\rho$ on $M$ follows naturally:
\begin{align} \label{entropy}
S(\rho,M) &= - \sum_\al \Tr_\al (p_\al \rho_{A_\al} \log (p_\al \rho_{A_\al}))\nonumber \\
&= -\sum_\al p_\al \log p_\al + \sum_\al p_\al S(\rho_{A_\al})
\end{align}

The second term in \eqref{entropy} is simply the average of the von Neumann entropy for each block diagonal normalised reduced density matrix $\rho_{A_\al}$ and has an operational interpretation as the \emph{distillable entanglement} in the limit of a large number of copies of the state \cite{ghosh2015entanglement, soni2016aspects}. In contrast, the first term in \eqref{entropy} is the classical Shannon entropy of mixing between the different blocks. 

Lastly, we define the algebraic relative entropy of two states $\rho$ and $\sigma$ on $M$ in terms of the modular Hamiltonian $K^\rho_M \equiv -\op_\al (\log (p_\al \rho_{A_\al}) \ot I_{\ovl A_\al})$ to be
\begin{align}\label{RelativeEntropy}
S(\rho|\sigma, M) = -S(\rho, M) + {\mathbb E}_\rho (K^\sigma_M).
\end{align}
As with the ordinary relative entropy, the algebraic relative entropy is nonnegative and is zero if and only if $\rho_M = \sigma_M$.

We're now ready to discuss quantum erasure correcting codes using von Neumann algebras.

\section{Fully algebraic quantum erasure correction} \label{QEC}
In this section, we present a version of quantum erasure correction which best describes the properties we know of AdS/CFT. The theorem presented here is a generalization of theorem 5.1 in \cite{harlow2017ryu} which alleviates the assumption that the full physical Hilbert space is factorizable. If the reader is not completely familiar with the subject of quantum error correction, we strongly encourage reading the logical progression of the argument in sections 3-5 of \cite{harlow2017ryu} before moving on to theorem \ref{thrm:conditions}.

\begin{thrm}\label{thrm:conditions}
Consider a finite-dimensional Hilbert space $\Hp$ on which we have a von Neumann algebra $N$, and a subspace $\Hc$ of $\Hp$ on which we have a von Neumann algebra $M$. Let $\ket{\wt{\al,ij}} = \ket{\wt{\al,i}}_{a_\al} \ot \ket{\wt{\al,j}}_{\ovl a_\al}$ be an orthonormal basis for $\Hc$ which is compatible with the decomposition $\Hc = \op_{\al} (\CH_{a_\al} \ot \CH_{\ovl{a}_\al})$ induced by $M$. Similarly, $N$ induces a direct sum decomposition $\Hp = \op_{\be} (\CH_{A_\be} \ot \CH_{\ovl{A}_\be})$ on $\Hp$. Let $\ket{\phi} = \frac{1}{\sqrt{|R|}} \sum_{\al,i j} \ket{\al,ij}_R \ket{\wt{\al,ij}}_{phys}$ where $R$ is an auxiliary system whose dimension is equal to that of $\Hc$. Then the following statements are equivalent:
\begin{itemize}
\item[(i)] We can decompose $\CH_{A_\be} = \op_\al (\CH_{\Ao} \ot \CH_{\At}) \op \CH_{A_3^\be}$ for all $\be$, such that for each $\al$, $|\Ao| = |a_\al|$ and $|\At| > 0$ for at least one $\be$. Then, there exists a unitary transformation $U \in N$ and sets of orthonormal states $\ket{\chi_{\al,j}} \in \op_\be \CH_{\At \ovl{A}_\be}$ such that 
\begin{align} \label{unitarymap}
\ket{\wt{\al,ij}} = U \left[\ket{\al,i}_{\Ao} \ot \ket{\chi_{\al,j}}_{\op_\be \At  \ovl{A}_\be}\right].
\end{align}
Here $\ket{\al,i}_{\Ao}$ is an orthonormal basis for $\CH_{\Ao}$.

\item[(ii)] For any operator $\wt{O} \in M$, there exists an operator $O \in N$ such that for any state $\ket{\wt{\psi}} \in \Hc$, we have
\begin{align}
O\ket{\wt{\psi}} &= \wt{O}\ket{\wt{\psi}}\nonumber \\
O^\dagger \ket{\wt{\psi}} &= \wt{O}^\dagger \ket{\wt{\psi}}.
\end{align}

\item[(iii)] For any operator $X' \in N'$, we have
\begin{align}
P_{code} X' P_{code} = \wt{X}' P_{code}
\end{align}
with $\wt{X}'$ some element of $M'$ and $P_{code}$ the projection onto $\Hc$. 

\item[(iv)] Let $\rho = \ket{\phi}\bra{\phi}$. For any operator $\wt{O} \in M$, we have
\begin{align}
[O_R, \rho_{RN'}] = 0,
\end{align}
where $O_R$ is defined as the unique operator on $\CH_R$ such that
\begin{align} \label{transpose}
O_R\ket{\phi} &= \wt{O}\ket{\phi}\nonumber \\
O_R^\dagger \ket{\phi} &= \wt{O}^\dagger \ket{\phi}.
\end{align}
Explicitly, $O_R$ acts with the same matrix elements on $R$ as $\wt{O}^T$ does on $\Hc$.
\end{itemize}
\end{thrm}

This theorem establishes the equivalence of a set of conditions that characterize the ability of a code subspace to recover a logical subalgebra $M$ from the erasure of a subalgebra $N'$ on the physical Hilbert space $\Hp$. Condition (i) asserts the existence of a unitary mapping between physical and logical states that allows the code subspace to recover the states in $M$ on $N$ by applying $U^\dagger$. Condition (ii) says that every logical operator in $M$ has a representation in $N$ that acts the same way on the code subspace. (iii) is the condition that operators acting on the erased subalgebra $N'$ do not disturb the information in the subalgebra $M$ of the code subspace, and finally, condition (iv) says that degrees of freedom in $R$ that are entangled with the logical subalgebra $M$ are uncorrelated with the erased subalgebra $N'$. The full proof of Theorem \ref{thrm:conditions} is given in Appendix \ref{proof}.

\begin{figure} 
\includegraphics[width=8.5 cm]{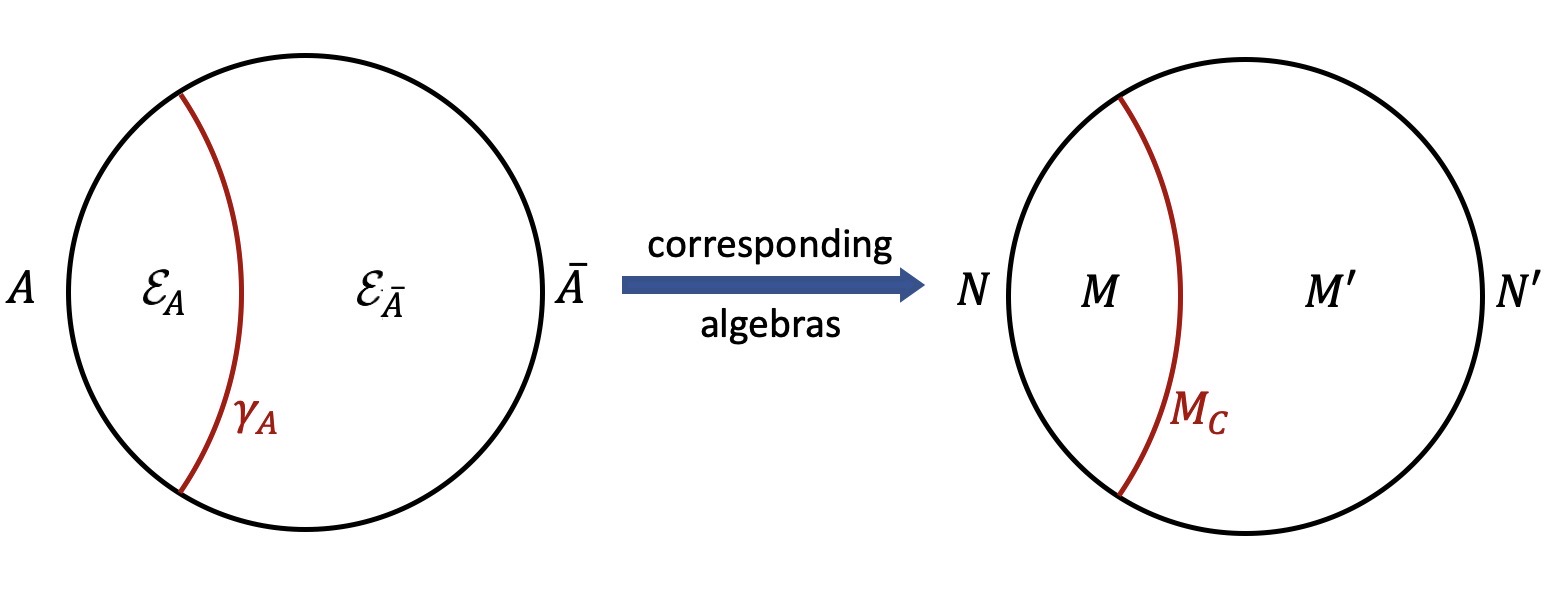}
\caption{An algebraic decomposition of the AdS/CFT geometry. On the left, we have a boundary subregion $A$ and its complement $\ovl A$, as well as their corresponding entanglement wedges ${\cal E}_A, {\cal E}_{\ovl A}$ in the bulk, separated by the RT surface $\gamma_A$. On the right, we see the von Neumann algebras $N, N', M, M',$ and $M_C$ that represent the degrees of freedom in $A, \ovl A, {\cal E}_A, {\cal E}_{\ovl A},$ and $\gamma_A$ respectively.} 
\label{fig:CorrespondingAlgebras}
\end{figure}

In the context of AdS/CFT, we can think of $\Hp$ as representing the boundary conformal field theory, and $\Hc$ as representing the low energy bulk effective field theory. Then $N$ represents the degrees of freedom in a boundary subregion, while $M$ represents the degrees of freedom in the entanglement wedge of that subregion, as illustrated in Figure \ref{fig:CorrespondingAlgebras}. Now condition (ii) of the theorem is nothing but a restatement of entanglement wedge reconstruction, i.e. that every operator in the entanglement wedge of a boundary subregion $A$ has a boundary representation with support only on $A$. 

AdS/CFT has the further property of complementary recovery which says that given a subregion $A$ on the boundary, not only can we represent operators in ${\cal E}_A$ as boundary operators on $A$, but we can also represent operators in ${\mathcal E}_{\ovl A}$ as boundary operators on $\ovl A$ (complement of $A$). Imposing this property of complementary recovery is equivalent to requiring, in addition to condition (ii), that any operator $\wt O' \in M'$ have a representation $O' \in N'$ that acts the same way on $\Hc$. We will call a code with such property a \textit{fully algebraic code with complementary recovery} (as opposed to Harlow's subalgebra code with complementary recovery). The equivalence of (ii) and (i) then implies that we must have
\begin{align}
U \ket{\al,i}_{\Ao} \ket{\chi_{\al,j}}_{\op_\be \At \ovl{A}_\be} = U' \ket{\al,j}_{\Aob} \ket{\chi_{\al,i}}_{\op_\be \Atb A_\be}
\end{align}
Here $U'$ is a unitary operator in $N'$ and we have decomposed $\CH_{\ovl A_\be} = \op_\al (\CH_{\Aob} \ot \CH_{\Atb}) \op \CH_{\ovl A_3^\be}$ for each $\be$ with $|\Aob| = |\ovl a_\al|$. $\ket{\al,j}_{\Aob}$ is an orthonormal basis for $\Aob$ and $\ket{\chi_{\al,i}}$ are orthonormal states in $\op_\be \CH_{\Atb A_\be}$. After some algebraic manipulation (see Appendix \ref{RTder}), we find that there must exist a state $\ket{\chi_\al} \in \op_\be \CH_{\At \Atb}$ such that
\begin{align} \label{CompRecUnitaryMap}
\ket{\wt {\al,ij}} = U U' \ket{\al,i}_{\Ao} \ket{\al,j}_{\Aob} \ket{\chi_\al}_{\op_{\be} \At \Atb}.
\end{align}
In the next section, we use this equation to derive an algebraic version of the Ryu-Takayanagi formula.

\section{Algebraic Ryu-Takayanagi formula and relative entropy} \label{RTfromQEC}
Given any state $\wt \rho$ in a fully algebraic code with complementary recovery on subalgebras $N$ and $N'$, we wish to find an expression for the algebraic entropy $S(\wt \rho, N)$, as well as $S(\wt \rho, N')$. In order to use the definition of algebraic entropy given in eq. \eqref{entropy}, we first need to find the diagonal blocks of $\wt \rho$ in $\be$. This can be accomplished by taking advantage of the encoding map in eq. \eqref{CompRecUnitaryMap}. We can express $\wt \rho$ as
\begin{align}
\wt \rho &= U' U [\op_\al (p_\al \rho_{\Ao\Aob} \ot (\chi_\al)_{\op_\be \At\Atb})] U^\dagger U'^\dagger
\end{align}
where $\rho_{\Ao\Aob}$ acts on $\CH_{\Ao} \ot \CH_{\Aob}$ the same way that $\wt \rho_{a_\al \ovl a_\al}$ acts on $\CH_{a_\al} \ot \CH_{\ovl a_\al}$ ($p_\al \rho_{a_\al \ovl a_\al}$ are the diagonal blocks of $\wt \rho$ in $\al$ with $p_\al$ chosen such that $\Tr\rho_{a_\al \ovl a_\al} = 1$), and we have defined the density matrix $\chi_\al = \ket{\chi_\al}\bra{\chi_\al}$. We see that the diagonal blocks of $\wt \rho$ in $\be$ are $$U'_{\ovl A_\be} U_{A_\be} [\op_\al (p_\al \rho_{\Ao \Aob} \ot k_{\al \be} (\chi_\al)_{\At \Atb})] U^\dagger_{A_\be} U'^\dagger_{\ovl A_\be}$$ where $k_{\al \be} $ is chosen so that $\Tr (\chi_\al)_{\At \Atb} = 1$. From eq. \eqref{rhoM} we then find
\begin{align}
\wt \rho_N &= U \left[ \op_\be \left( [ \op_\al (p_\al \rho_{\Ao} \ot k_{\al\be} ({\chi_\al})_{\At}) ] \ot \frac{I_{{\ovl A}_\be}}{{|\ovl A}_\be|} \right) \right] U^\dagger\\
\wt \rho_{N'} &= U' \left[ \op_\be \left( \frac{I_{A_\be}}{|A_\be|} \ot [ \op_\al (p_\al \rho_{\Aob} \ot k_{\al\be} ({\chi_\al})_{\Atb}) ] \right) \right] U'^\dagger
\end{align}
We can now use the definition given in eq. \eqref{entropy} to derive the algebraic entropies (Appendix \ref{RTder}) and show that they satisfy
\begin{align}
S(\wt \rho, N) &= \Tr (\wt \rho {\cal L}) + S(\wt \rho, M) \label{AlgRT}\\
S(\wt \rho, N') &= \Tr (\wt \rho {\cal L}) + S(\wt \rho, M') \label{AlgRTComp}
\end{align}
where we have defined ${\cal L} = \op_\al S(\chi_\al, N) I_{a_\al {\ovl a}_\al}$. We see that eqns. \eqref{AlgRT} and \eqref{AlgRTComp} are in compliance with the RT formula given in eq. \eqref{RT} and ${\cal L}$ is the equivalent of the area operator. This area operator has the desired properties we expect from holography. First of all, we note that it lies in the center $M_C$ of $M$, which is by definition the intersection of $M$ and $M'$, i.e. the algebra that corresponds to the RT surface. Secondly, $S(\chi_\al, N)$ can have a different value for each $\al$, so the area operator can be nontrivial. Thirdly, we see that ${\cal L}$ is independent of $\wt \rho$, but depends on $N$ and $M$ as we would expect. 

We also note that the ``areas'' $S(\chi_\al, N)$ naturally decompose into two pieces, a distillable part
$$
\sum_\beta k_{\alpha \beta} S(({\chi_\al})_{A_2^{\alpha \beta}}),
$$
and a Shannon entropy from the mixture of boundary superselection sectors
$$
-\sum_\beta k_{\alpha \beta} \log k_{\alpha \beta}.
$$

Conversely, one can also show that any pair of bulk and boundary algebras $M,N$, satisfying \eqref{AlgRT} and \eqref{AlgRTComp}, have complementary recovery. The full proof is again given in Appendix \ref{RTder}, but the basic strategy is to show that \eqref{AlgRT} and \eqref{AlgRTComp} imply that the "bulk" and "boundary" relative entropies are equal. Specifically,
\begin{align}
S(\wt \rho|\wt \sigma, N) &= S(\wt \rho|\wt \sigma, M)\\
S(\wt \rho|\wt \sigma, N') &= S(\wt \rho|\wt \sigma, M'),
\end{align}
where the algebraic relative entropy $S(\wt \rho|\wt \sigma, M)$ was derived in \eqref{RelativeEntropy}. This is a fully algebraic version of the results of \cite{jafferis2016relative}. From this one can easily prove condition (iii) of Theorem \ref{thrm:conditions}.

\section{Other boundary entropies} \label{OtherEntropies}
The algebraic entropy is not the only definition of entanglement entropy for gauge theories that has been considered in the literature.

Another candidate entropy is the distillable entropy \cite{ghosh2015entanglement, soni2016aspects}, which differs from the algebraic entropy by the absence of a classical Shannon term from the mixture of superselection sectors. In other words, we have
\begin{align}
S_\text{distill}(\rho , N) = \sum_\beta p_\beta S(\rho_{A_\beta}).
\end{align}
This entropy reduces to the von Neumann entropy in the case of only one superselection sector $\beta$, just like the algebraic entropy. It also preserves an important operational property of the von Neumann entropy. Specifically, it agrees with the rate at which one can extract Bell pairs, from a large number of copies of the state, by local operators in the algebra and commutant, as well as classical communication.

Since the distillable entropy differs from the algebraic entropy by the nonlinear Shannon entropy term, it does not obey eq. \eqref{AlgRT}. This is still true, even if  the algebraic bulk entropy is replaced by the distillable bulk entropy, because the formula needs to hold when the boundary theory factorises. In that case, the boundary algebraic and distillable entropies agree, but the bulk algebraic and distllable entropies differ by a nonzero and nonlinear term.

Finally, there is a third natural definition of the boundary entropy, which we call the extended Hilbert space entropy. It is also sometimes called the ``full'' entropy or the ``$\log \text{dim} R$'' entropy \cite{donnelly2012decomposition, donnelly2014entanglement, lin2018comments}. This is defined as the von Neumann entropy when the physical Hilbert space is embedded in a larger, factorised Hilbert space, such that the algebra and its commutant can be reconstructed on different factors of the larger extended Hilbert space.

This extended Hilbert space is in general nonunique. Indeed the simplest such Hilbert space is
\begin{align}
    \CH = (\oplus_\beta {\CH}_{A_\beta}) \otimes (\oplus_{\beta'} \CH_{\bar{A}_{\beta'}}),
\end{align}
in which case the extended Hilbert space entropy will be equal to the algebraic entropy. However, for non-Abelian lattice gauge theories, there is a particularly natural choice of extended Hilbert space, namely the Hilbert space found by not imposing gauge constraints at the boundary of the two regions. This gives an entanglement entropy
\begin{align}
S_\text{extend}(\rho , N) = S(\rho, N) + \sum_e \sum_R p_{e,R} \log \text{dim } R.
\end{align}
where the second term involves a sum over all edges $e$ at the boundary of the two regions and irreducible representations $R$ of the gauge group and $p_{e,R}$ is the sum over probabilities $p_\beta$ for all superselection sectors where edge $e$ is in representation $R$. This second term is linear and can be thought of as an additional ``area term'' in a Ryu-Takayanagi formula for the encoding of the gauge-invariant algebras in the extended Hilbert space.

Since the extended Hilbert space entropy differs from the algebraic entropy by a linear term, it is perfectly possible that both satisfy a version of the Ryu-Takayanagi formula, with different area operators $\cal L$. Indeed, it is easy to see that this is the case: since the bulk code space is a subspace of the extended Hilbert space, the extended Hilbert space factorises, and we have complementary recovery, a Ryu-Takayanagi formula for the extended Hilbert space entropy follows immediately from the results of \cite{harlow2017ryu}.

An obvious question is which entropy obeys the 'real' Ryu-Takayanagi formula, where $\cal L$ is the actual bulk area. However, the two entropies only differ by a sum of local boundary observables. Such observables cannot know about physics deep in the bulk; instead they can only be related to the details of the regularisation of the bulk area at the lattice scale. In particular, if we regulate the entropy at a lengthscale that is much larger than the lattice scale, for example by looking at the mutual information of two regions separated by a small distance (that is nonetheless much larger than the lattice scale), both entropies will give the same result.

The choice of which entropy to prefer is therefore something of a philosophical question. Fans of the extended Hilbert space entropy will emphasize its tighter connection to path integral replica trick calculations (since path integrals are generally done in the extended Hilbert space) \cite{donnelly2012decomposition, donnelly2014entanglement}. We tend to prefer the algebraic entropy, since it does not rely on an unphysical, and somewhat arbitrary, choice of extended Hilbert space. It also seems more closely connected to continuum definitions of relative entropy, and in particular mutual information, via Tomita-Takesaki theory \cite{witten2018notes}.

\section{Discussion of results} \label{discusson}
We have characterized a fully algebraic code that describes the error correcting properties of the AdS/CFT bulk to boundary map in the typical case where the conformal field theory has a gauge symmetry and the boundary Hilbert space does not necessarily factorise. We have also shown that when this code has a complementary recovery property similar to that of AdS/CFT, it yields a version of the Ryu-Takayanagi formula and an equation between the bulk and boundary relative entropies. We present the summary of our main results as a theorem that, when applied to AdS/CFT, establishes an equivalence between entanglement wedge reconstruction and the Ryu-Takayanagi formula. This theorem is the equivalent of Theorem 1.1 from \cite{harlow2017ryu}.

\begin{thrm} \label{main}
Consider a finite-dimensional Hilbert space $\Hp$ and a subspace $\Hc \subset \Hp$. Let $N$ and $M$ be von Neumann algebras on $\Hp$ and $\Hc$ respectively. Then the following statements are equivalent:
\begin{itemize}
\item[(a)] For any operators $\wt O \in M, \wt O' \in M'$, there exist operators $O \in N$ and $O' \in N'$ such that for any state $\ket{\wt{\psi}} \in \Hc$, we have 
\begin{align} \nonumber
O\ket{\wt{\psi}} = \wt{O}\ket{\wt{\psi}} \hspace{4pt} &,\hspace{4pt}
O^\dagger \ket{\wt{\psi}} = \wt{O}^\dagger \ket{\wt{\psi}} \\ \nonumber
O'\ket{\wt{\psi}} = \wt O' \ket{\wt{\psi}} \hspace{4pt} &, \hspace{4pt}
O'^\dagger \ket{\wt{\psi}} = \wt O'^\dagger \ket{\wt{\psi}}.
\end{align} 
\item[(b)] There exists an operator ${\cal L} \in M_C$ such that for any state $\wt \rho$ in $\Hc$, we have
\begin{align} \nonumber
S(\wt \rho, N) &= \Tr (\wt \rho {\cal L}) + S(\wt \rho, M) \\ \nonumber
S(\wt \rho, N') &= \Tr (\wt \rho {\cal L}) + S(\wt \rho, M').
\end{align}
\item[(c)] For any state $\wt \rho$ and $\wt \sigma$ in $\Hc$, we have
\begin{align} \nonumber
S(\wt \rho| \wt \sigma, N) &= S(\wt \rho| \wt \sigma, M)\\ \nonumber
S(\wt \rho| \wt \sigma, N') &= S(\wt \rho| \wt \sigma, M')
\end{align}
\end{itemize}
\end{thrm}

The Ryu-Takayanagi formula presented in part (b) of Theorem \ref{main} has a nice symmetry compared to the one obtained in \cite{harlow2017ryu} as the same definition of entropy is used throughout the formula. In fact, as we showed in Section \ref{RTfromQEC}, this algebraic entropy also shows up in the definition of the area operator $\cal L$. As a result, we find that $\cal L$ naturally decomposes into two parts. One part comes from entanglement within a single boundary superselection sector, while the other comes from a classical Shannon entropy added by the mixture between superselection sectors. We leave the question of whether both terms actually contribute to the classical area at leading order in large $N$ holographic theories to future work.

We wish to emphasize that the results presented here represent an idealised version of the true correspondence between bulk and boundary algebras in AdS/CFT. We have ignored errors in the reconstruction of the bulk algebras \cite{cotler2019entanglement, chen2019entanglement} and corrections to the Ryu-Takayanagi formula \cite{dong2014holographic}, as well as the fact that complementary recovery breaks down at higher orders in $G_N$ \cite{dong2014holographic}. We have also treated both the bulk and boundary algebras as finite-dimensional. For the boundary algebras, one should really instead use the full continuum CFT algebras, which are infinite-dimensional type III von Neumann algebras. However, this adds considerable complications: for instance, entropies need to be regulated (e.g. using the mutual information) to be well defined. See \cite{kang2018holographic} for a proof of the equivalence of equalities between bulk and boundary relative entropies and entanglement wedge reconstruction in this more general setting. We are somewhat more skeptical that allowing the bulk algebra to be an infinite-dimensional von Neumann algebra, as in \cite{kang2018holographic}, makes results more physically realistic, although it certainly doesn't hurt, since the bulk effective field theory description will break down anyway if we try to add too much entropy in a small region.

\section{Acknowledgements}
We would like to thank Isaac Kim, Sepehr Nezami, Grant Salton, Jon Sorce, and Michael Walter for detailed comments and fruitful discussions. In particular, we would like to thank Patrick Hayden for his invaluable support and mentorship. GP is supported in part by AFOSR award FA9550-16-1- 0082 and DOE award {DE-SC0019380}.

\bibliographystyle{IEEEtran}
\bibliography{AlgebraicRT}

\begin{thebibliography}{10}
\providecommand{\url}[1]{#1}
\csname url@samestyle\endcsname
\providecommand{\newblock}{\relax}
\providecommand{\bibinfo}[2]{#2}
\providecommand{\BIBentrySTDinterwordspacing}{\spaceskip=0pt\relax}
\providecommand{\BIBentryALTinterwordstretchfactor}{4}
\providecommand{\BIBentryALTinterwordspacing}{\spaceskip=\fontdimen2\font plus
\BIBentryALTinterwordstretchfactor\fontdimen3\font minus
  \fontdimen4\font\relax}
\providecommand{\BIBforeignlanguage}[2]{{%
\expandafter\ifx\csname l@#1\endcsname\relax
\typeout{** WARNING: IEEEtran.bst: No hyphenation pattern has been}%
\typeout{** loaded for the language `#1'. Using the pattern for}%
\typeout{** the default language instead.}%
\else
\language=\csname l@#1\endcsname
\fi
#2}}
\providecommand{\BIBdecl}{\relax}
\BIBdecl

\bibitem{maldacena1999large}
J.~Maldacena, ``The large-{N} limit of superconformal field theories and
  supergravity,'' \emph{International journal of theoretical physics}, vol.~38,
  no.~4, pp. 1113--1133, 1999.

\bibitem{witten1998anti}
E.~Witten, ``Anti de {Sitter} space and holography,'' \emph{arXiv preprint
  hep-th/9802150}, 1998.

\bibitem{ryu2006holographic}
S.~Ryu and T.~Takayanagi, ``Holographic derivation of entanglement entropy from
  the anti--de {Sitter} space/conformal field theory correspondence,''
  \emph{Physical review letters}, vol.~96, no.~18, p. 181602, 2006.

\bibitem{ryu2006aspects}
------, ``Aspects of holographic entanglement entropy,'' \emph{Journal of High
  Energy Physics}, vol. 2006, no.~08, p. 045, 2006.

\bibitem{almheiri2015bulk}
A.~Almheiri, X.~Dong, and D.~Harlow, ``Bulk locality and quantum error
  correction in ads/cft,'' \emph{Journal of High Energy Physics}, vol. 2015,
  no.~4, p. 163, 2015.

\bibitem{mintun2015bulk}
E.~Mintun, J.~Polchinski, and V.~Rosenhaus, ``Bulk-boundary duality, gauge
  invariance, and quantum error corrections,'' \emph{Physical review letters},
  vol. 115, no.~15, p. 151601, 2015.

\bibitem{pastawski2015holographic}
F.~Pastawski, B.~Yoshida, D.~Harlow, and J.~Preskill, ``Holographic quantum
  error-correcting codes: Toy models for the bulk/boundary correspondence,''
  \emph{Journal of High Energy Physics}, vol. 2015, no.~6, p. 149, 2015.

\bibitem{hayden2016holographic}
P.~Hayden, S.~Nezami, X.-L. Qi, N.~Thomas, M.~Walter, and Z.~Yang,
  ``Holographic duality from random tensor networks,'' \emph{Journal of High
  Energy Physics}, vol. 2016, no.~11, p.~9, 2016.

\bibitem{verlinde2017emergent}
E.~P. Verlinde, ``Emergent gravity and the dark universe,'' \emph{SciPost
  Phys}, vol.~2, no.~3, p. 016, 2017.

\bibitem{hayden2018learning}
P.~Hayden and G.~Penington, ``Learning the alpha-bits of black holes,''
  \emph{arXiv preprint arXiv:1807.06041}, 2018.

\bibitem{penington2019entanglement}
G.~Penington, ``Entanglement wedge reconstruction and the information
  paradox,'' \emph{arXiv preprint arXiv:1905.08255}, 2019.

\bibitem{almheiri2019entropy}
A.~Almheiri, N.~Engelhardt, D.~Marolf, and H.~Maxfield, ``The entropy of bulk
  quantum fields and the entanglement wedge of an evaporating black hole,''
  \emph{arXiv preprint arXiv:1905.08762}, 2019.

\bibitem{harlow2017ryu}
D.~Harlow, ``The ryu--takayanagi formula from quantum error correction,''
  \emph{Communications in Mathematical Physics}, vol. 354, no.~3, pp. 865--912,
  2017.

\bibitem{faulkner2013quantum}
T.~Faulkner, A.~Lewkowycz, and J.~Maldacena, ``Quantum corrections to
  holographic entanglement entropy,'' \emph{Journal of High Energy Physics},
  vol. 2013, no.~11, p.~74, 2013.

\bibitem{czech2012gravity}
B.~Czech, J.~L. Karczmarek, F.~Nogueira, and M.~Van~Raamsdonk, ``The gravity
  dual of a density matrix,'' \emph{Classical and Quantum Gravity}, vol.~29,
  no.~15, p. 155009, 2012.

\bibitem{wall2014maximin}
A.~C. Wall, ``Maximin surfaces, and the strong subadditivity of the covariant
  holographic entanglement entropy,'' \emph{Classical and Quantum Gravity},
  vol.~31, no.~22, p. 225007, 2014.

\bibitem{headrick2014causality}
M.~Headrick, V.~E. Hubeny, A.~Lawrence, and M.~Rangamani, ``Causality \&
  holographic entanglement entropy,'' \emph{Journal of High Energy Physics},
  vol. 2014, no.~12, p. 162, 2014.

\bibitem{dong2016reconstruction}
X.~Dong, D.~Harlow, and A.~C. Wall, ``Reconstruction of bulk operators within
  the entanglement wedge in gauge-gravity duality,'' \emph{Physical review
  letters}, vol. 117, no.~2, p. 021601, 2016.

\bibitem{jafferis2016relative}
D.~L. Jafferis, A.~Lewkowycz, J.~Maldacena, and S.~J. Suh, ``Relative entropy
  equals bulk relative entropy,'' \emph{Journal of High Energy Physics}, vol.
  2016, no.~6, p.~4, 2016.

\bibitem{faulkner2017bulk}
T.~Faulkner and A.~Lewkowycz, ``Bulk locality from modular flow,''
  \emph{Journal of High Energy Physics}, vol. 2017, no.~7, p. 151, 2017.

\bibitem{cotler2019entanglement}
J.~Cotler, P.~Hayden, G.~Penington, G.~Salton, B.~Swingle, and M.~Walter,
  ``Entanglement wedge reconstruction via universal recovery channels,''
  \emph{Physical Review X}, vol.~9, no.~3, p. 031011, 2019.

\bibitem{witten2018notes}
E.~Witten, ``Notes on some entanglement properties of quantum field theory,''
  \emph{arXiv preprint arXiv:1803.04993}, 2018.

\bibitem{casini2014remarks}
H.~Casini, M.~Huerta, and J.~A. Rosabal, ``Remarks on entanglement entropy for
  gauge fields,'' \emph{Physical Review D}, vol.~89, no.~8, p. 085012, 2014.

\bibitem{kang2018holographic}
M.~J. Kang and D.~K. Kolchmeyer, ``Holographic relative entropy in
  infinite-dimensional hilbert spaces,'' \emph{arXiv preprint
  arXiv:1811.05482}, 2018.

\bibitem{ghosh2015entanglement}
S.~Ghosh, R.~M. Soni, and S.~P. Trivedi, ``On the entanglement entropy for
  gauge theories,'' \emph{Journal of High Energy Physics}, vol. 2015, no.~9,
  p.~69, 2015.

\bibitem{soni2016aspects}
R.~M. Soni and S.~P. Trivedi, ``Aspects of entanglement entropy for gauge
  theories,'' \emph{Journal of High Energy Physics}, vol. 2016, no.~1, p. 136,
  2016.

\bibitem{donnelly2012decomposition}
W.~Donnelly, ``Decomposition of entanglement entropy in lattice gauge theory,''
  \emph{Physical Review D}, vol.~85, no.~8, p. 085004, 2012.

\bibitem{donnelly2014entanglement}
------, ``Entanglement entropy and nonabelian gauge symmetry,'' \emph{Classical
  and Quantum Gravity}, vol.~31, no.~21, p. 214003, 2014.

\bibitem{lin2018comments}
J.~Lin and D.~Radicevic, ``Comments on defining entanglement entropy,''
  \emph{arXiv preprint arXiv:1808.05939}, 2018.

\bibitem{hubeny2007covariant}
V.~E. Hubeny, M.~Rangamani, and T.~Takayanagi, ``A covariant holographic
  entanglement entropy proposal,'' \emph{Journal of High Energy Physics}, vol.
  2007, no.~07, p. 062, 2007.

\bibitem{engelhardt2015quantum}
N.~Engelhardt and A.~C. Wall, ``Quantum extremal surfaces: holographic
  entanglement entropy beyond the classical regime,'' \emph{Journal of High
  Energy Physics}, vol. 2015, no.~1, p.~73, 2015.

\bibitem{dong2018entropy}
X.~Dong and A.~Lewkowycz, ``Entropy, extremality, euclidean variations, and the
  equations of motion,'' \emph{Journal of High Energy Physics}, vol. 2018,
  no.~1, p.~81, 2018.

\bibitem{chen2019entanglement}
C.~F. Chen, G.~Penington, and G.~Salton, ``Entanglement wedge reconstruction
  using the petz map,'' \emph{arXiv preprint arXiv:1902.02844}, 2019.

\bibitem{dong2014holographic}
X.~Dong, ``Holographic entanglement entropy for general higher derivative
  gravity,'' \emph{Journal of High Energy Physics}, vol. 2014, no.~1, p.~44,
  2014.

\end{thebibliography}
\clearpage

\appendix
\begin{widetext}

\section{Proof of Theorem \ref{thrm:conditions}} \label{proof}

$(i) \implies (ii)$: $\wt O$ is in $M$ and therefore has the form $\wt O = \op_\al (\wt O_{a_\al} \ot I_{\ovl a_\al})$. We can then simply define $O \equiv U \left(\op_{\al, \be} (O_{\Ao} \ot I_{\At \ovl A_\be})\right) U^\dagger$, where $O_{\Ao}$ acts on $\CH_{\Ao}$ in the same way that $\wt{O}_{a_\al}$ acts on $\CH_{a_\al}$.\\

 $(ii) \implies(iii)$: We will give a proof by contradiction. Let $X'$ be an operator in $N'$ and assume $P_{code} X' P_{code} = \wt{x} P_{code}$ with $\wt{x}$ an operator on $\Hc$, but not an element of $M'$. By definition of the commutant then there must exist an operator $\wt{O} \in M$ that $\wt{x}$ doesn't commute with. Thus, there exists a state $\ket{\wt{\psi}} \in \Hc$ such that $\bra{\wt{\psi}}[\wt{x}, \wt{O}]\ket{\wt{\psi}} \neq 0$. Now $(ii)$ implies that there exists an operator $O \in N$ such that $\bra{\wt{\psi}}[\wt{x}, \wt{O}]\ket{\wt{\psi}} = \bra{\wt{\psi}}[\wt{x}, O]\ket{\wt{\psi}} = \bra{\wt{\psi}}[X', O]\ket{\wt{\psi}} \neq 0$, but this is a contradiction since by definition, $X'$ commutes with all operators in $N$.\\

$(iii) \implies (iv)$: For any operator $X' \in N'$ and $Y_R$ acting on $\CH_R$ and any operator $\wt{O} \in M$, we have $\Tr(O_R \rho_{RN'} X' Y_R) = \Tr(\rho_{RN'} X' Y_R O_R) = \Tr(\rho X' Y_R O_R) = \bra{\phi} X' Y_R O_R \ket{\phi} = \bra{\phi} \wt{X}' Y_R \wt{O} \ket{\phi} = \bra{\phi} \wt{O} \wt{X}' Y_R \ket{\phi} = \bra{\phi} O_R X' Y_R \ket{\phi} = \Tr(\rho O_R X' Y_R) = \Tr( \rho_{RN'}O_R X' Y_R)$ where $\wt{X}'$ and $O_R$ are defined as in $(iii)$ and $(iv)$, respectively. We just showed that $\Tr([O_R, \rho_{RN'}] X' Y_R) = 0$, which can only be true for arbitrary $X'$ and $Y_R$ if $[O_R, \rho_{RN'}] = 0$.\\

$(iv) \implies (i)$: We can write $\ket{\phi} = \op_\be c_\be \ket{\phi_\be}$ in accordance with the Hilbert space decomposition $\Hp = \op_{\be} (\CH_{A_\be} \ot \CH_{\ovl{A}_\be})$ such that $\{\ket{\phi_\be}\}$ is a set of orthonormal states and $\sum_\be |c_\be|^2 = 1$. Then the diagonal blocks of $\rho$ in $\be$ look like  $|c_\be|^2 \rho_{R A_\be \ovl A_\be}$ where $\rho_{R A_\be \ovl A_\be} = \ket{\phi_\be}\bra{\phi_\be}$ has unit trace. Thus, we'll have
\begin{align}\nonumber
\rho_{R N'} = \op_\be \left(\frac{I_{A_\be}}{|A_\be|} \ot |c_\be|^2 \rho_{R \ovl{A}_\be} \right).
\end{align}
Now note that the basis $\ket{\al, ij}_R$ for $\CH_R$ gives the decomposition
\begin{align}
\CH_{R \ovl{A}_\be} = \left(\op_\al (\CH_{R_\al} \ot \CH_{\ovl{R}_\al}) \right) \ot \CH_{\ovl{A}_\be} = \op_\al (\CH_{R_\al} \ot \CH_{\ovl{R}_\al} \ot \CH_{\ovl{A}_\be})
\end{align}
under which $(iv)$ implies that
\begin{align}
\rho_{R\ovl{A}_\be} = \op_\al \left[ \frac{| R_\al| |\ovl{R}_\al|}{|R|} \left(\frac{I_{R_\al}}{|R_\al|} \ot \rho_{\ovl{R}_\al\ovl{A}_\be} \right) \right]
\end{align}
for some $\rho_{\ovl{R}_\al\ovl{A}_\be}$, where we have used the fact that $\rho_R = \frac{I_R}{|R|}$ to determine the coefficients. Since $\ket{\phi_\be}$ purifies $\rho_{R\ovl{A}_\be}$, by the Schmidt decomposition, the rank of $\rho_{R\ovl A_\be}$ must be less than $|A_\be|$, that is,
\begin{align}
\sum_\al |R_\al||\rho_{\ovl{R}_\al\ovl{A}_\be}| \leq |A_\be|
\end{align}
where $|\rho_{\ovl{R}_\al\ovl{A}_\be}|$ denotes the rank of $\rho_{\ovl{R}_\al\ovl{A}_\be}$. Therefore we can decompose
\begin{align}
\CH_{A_\be} = \op_\al \left(\CH_{A_1^\al} \ot \CH_{\At}\right) \op \CH_{A_3^\be}
\end{align}
such that $|A_1^\al| = |R_\al| = |a_\al|$ and $|\At| \geq |\rho_{\ovl{R}_\al\ovl{A}_\be}|$. This implies that $\rho_{\ovl{R}_\al\ovl{A}_\be}$ has a purification on $\At$, which we will denote by $\ket{\psi_{\al,\be}}_{\ovl R_\al \At \ovl A_\be}$. We can then use this to purify $\rho_{R\ovl{A}_\be}$ on $A_\be$ as
\begin{align}
\ket{\phi'_\be} = \frac{1}{\sqrt{|R|}} \sum_{\al, i} \sqrt{|\ovl R_\al|} \ket{\al, i}_{R_\al} \ket{\al, i}_{\Ao} \ket{\psi_{\al,\be}}_{\ovl R_\al \At \ovl A_\be}
\end{align}
Since $\ket{\phi_\be}$ and $\ket{\phi'_\be}$ are two purifications of $\rho_{R\ovl{A}_\be}$ on $A_\be$, they must differ only by a unitary $U_{A_\be}$. This then gives
\begin{align}\label{TwoPurifications} \nonumber
\ket{\phi} &= \sum_\be c_\be \ket{\phi_\be}\\\nonumber
&= \sum_\be c_\be U_{A_\be}\ket{\phi'_\be}\\
&= U \left( \frac{1}{\sqrt{|R|}} \sum_{\al, i} \sqrt{|\ovl R_\al|}  \ket{\al, i}_{R_\al} \ket{\al, i}_{\Ao} \sum_\be c_\be  \ket{\psi_{\al,\be}}_{\ovl R_\al \At \ovl A_\be} \right)
\end{align}
where $U = \op_\be (U_{A_\be} \ot I_{\ovl{A}_\be}) \in N$. We know from $\rho_R =  \frac{I_R}{|R|}$ that we must have $\sum_\be  |c_\be|^2 \Tr_{\ovl A_\be} \rho_{\ovl R_\al \ovl A_\be} = \sum_\be  |c_\be|^2 \Tr_{\At \ovl A_\be}\ket{\psi_{\al,\be}}\bra{\psi_{\al,\be}} = \Tr_{phys}(\op_\be |c_\be|^2 \ket{\psi_{\al,\be}}\bra{\psi_{\al,\be}}) =  \frac{I_{\ovl R_\al}}{|\ovl R_\al|}$, which means the state $\sum_\be c_\be  \ket{\psi_{\al,\be}}$ on $\ovl R_\al (\op_\be \At \ovl A_\be)$  must have the form
\begin{align}\label{MaximallyEntangled}
\sum_\be c_\be  \ket{\psi_{\al,\be}}_{\ovl R_\al \At \ovl A_\be} = \frac{1}{\sqrt{|\ovl R_\al|}} \sum_j \ket{\al, j}_{\ovl R_\al} \ket{\chi_{\al,j}}_{\op_\be {\At \ovl{A}_\be}}
\end{align}
for some orthonormal $\ket{\chi_{\al,j}}_{\op_\be {\At \ovl{A}_\be}}$. Combining \eqref{MaximallyEntangled} with \eqref{TwoPurifications} gives
\begin{align}
\sum_{\al,i j} \ket{\al,ij}_R \ket{\wt{\al,ij}}_{phys} = U \left( \sum_{\al, ij}   \ket{\al, ij}_R \ket{\al, i}_{\Ao} \ket{\chi_{\al,j}}_{\op_\be {\At \ovl{A}_\be}} \right),
\end{align}
which implies $(i)$.

\section{Ryu-Takayanagi formula from a fully algebraic code with complementary recovery} \label{RTder}
Consider a fully algebraic code as described in Theorem \ref{thrm:conditions} such that for every operator $\wt O \in M$, there exists an operator $O \in N$ that acts the same way on $\Hc$ (condition (ii)). Moreover, assume that for every operator $\wt O' \in M'$, there exists an operator $O' \in N'$ that acts the same way on $\Hc$. The equivalence of (ii) and (i) in Theorem \ref{thrm:conditions} then implies that there exist unitary transformations $U \in N$ and $U' \in N'$, and sets of orthonormal states $\ket{\chi_{\al,j}} \in \op_\be \CH_{\At \ovl{A}_\be}$ and $\ket{\ovl \chi_{\al,i}} \in \op_\be \CH_{\Atb A_\be}$ such that
\begin{align}
\ket{\al, ij} = U \ket{\al,i}_{\Ao} \ket{\chi_{\al,j}}_{\op_\be \At \ovl{A}_\be} = U' \ket{\al,j}_{\Aob} \ket{\ovl \chi_{\al,i}}_{\op_\be \Atb A_\be}.
\end{align}
Multiplying the above equation by $U^\dagger U'^\dagger$ we get
\begin{align} \label{ijeq}
\ket{\al,i}_{\Ao} U'^\dagger \ket{\chi_{\al,j}}_{\op_\be \At \ovl{A}_\be} = \ket{\al,j}_{\Aob} U^\dagger \ket{\ovl \chi_{\al,i}}_{\op_\be \Atb A_\be},
\end{align}
which implies there must be states $\ket{\chi_\al}_{\op_\be \At \Atb}$ and  $\ket{\ovl \chi_\al}_{\op_\be \Atb \At}$ such that
\begin{align}
U'^\dagger  \ket{\chi_{\al,j}}_{\op_\be \At \ovl{A}_\be} &= \ket{\al,j}_{\Aob} \ket{\chi_\al}_{\op_\be \At \Atb} \\
U^\dagger \ket{\ovl \chi_{\al,i}}_{\op_\be \Atb A_\be} &= \ket{\al,i}_{\Ao} \ket{\ovl \chi_\al}_{\op_\be \Atb \At}.
\end{align}
Plugging these expressions back into \eqref{ijeq} tells us that in fact $\ket{\chi_\al}_{\op_\be \At \Atb} = \ket{\ovl \chi_\al}_{\op_\be \Atb \At}$. Therefore we obtain
\begin{align} \label{CompRecUnitaryMapAppendix}
\ket{\wt {\al,ij}} = U U' \ket{\al,i}_{\Ao} \ket{\al,j}_{\Aob} \ket{\chi_\al}_{\op_{\be} \At \Atb}.
\end{align}

The encoding map in eq. \eqref{CompRecUnitaryMapAppendix} describes a code with complementary recovery on subalgebras $N$ and $N'$. For any state $\wt \rho$ in the code space, we can use this map to express $\wt \rho$ as
\begin{align}\nonumber
\wt \rho &= U' U (\op_\al (p_\al \rho_{\Ao\Aob} \ot (\chi_\al)_{\op_\be \At\Atb})) U^\dagger U'^\dagger
\end{align}
where $\rho_{\Ao\Aob}$ acts on $\CH_{\Ao} \ot \CH_{\Aob}$ the same way that $\wt \rho_{a_\al \ovl a_\al}$ acts on $\CH_{a_\al} \ot \CH_{\ovl a_\al}$ ($p_\al \rho_{a_\al \ovl a_\al}$ are the diagonal blocks of $\wt \rho$ in $\al$ with $p_\al$ chosen such that $\Tr\rho_{a_\al \ovl a_\al} = 1$), and we have defined the density matrix $\chi_\al = \ket{\chi_\al}\bra{\chi_\al}$. We see that the diagonal blocks of $\wt \rho$ in $\be$ are $$U'_{\ovl A_\be} U_{A_\be} [\op_\al (p_\al \rho_{\Ao \Aob} \ot k_{\al \be} (\chi_\al)_{\At \Atb})] U^\dagger_{A_\be} U'^\dagger_{\ovl A_\be}$$ where $k_{\al \be} $ is chosen so that $\Tr (\chi_\al)_{\At \Atb} = 1$. This implies that $\sum_\be k_{\al \be} = 1$. We can then compute
\begin{align}
\wt \rho_N = U \left[ \op_\be \left( [ \op_\al (p_\al \rho_{\Ao} \ot k_{\al\be} ({\chi_\al})_{\At}) ] \ot \frac{I_{{\ovl A}_\be}}{{|\ovl A}_\be|} \right) \right] U^\dagger
\end{align}

\begin{align}\nonumber
S(\wt \rho, N) &= - \sum_\be \Tr_{A_\be} \left( [\op_\al(p_\al \rho_{\Ao} \ot k_{\al\be} ({\chi_\al})_{\At})] \log [\op_\al (p_\al \rho_{\Ao} \ot k_{\al\be} ({\chi_\al})_{\At})] \right)\\\nonumber
&= - \sum_{\al\be} \Tr_{A_\be} \left(p_\al \rho_{\Ao} \log (p_\al \rho_{\Ao}) \ot k_{\al\be} ({\chi_\al})_{\At} +  p_\al \rho_{\Ao}  \ot k_{\al\be} ({\chi_\al})_{\At} \log(k_{\al\be} ({\chi_\al})_{\At}) \right)\\\nonumber
&= - \sum_\al ( \Tr_{a_\al} p_\al \wt \rho_{a_\al} \log (p_\al \wt \rho_{a_\al})) ( \sum_\be k_{\al\be} ) + \sum_\al p_\al (- \sum_\be \Tr_{\At} k_{\al\be} ({\chi_\al})_{\At} \log(k_{\al\be} ({\chi_\al})_{\At}))\\\nonumber
&= S(\wt \rho, M) + \sum_\al p_\al S(\chi_\al, N)\\
&= S(\wt \rho, M) + \Tr (\wt \rho \cal L) \label{AlgRTAppendix}
\end{align}
where we have defined ${\cal L} = \op_\al S(\chi_\al, N) I_{a_\al {\ovl a}_\al}$. Here, we used the property that $S(\rho, N)$ is invariant under $\rho \rightarrow U\rho U^\dagger$ for any $U \in N$. The second equality follows from the identity $\log (O_A \ot O_B) = \log O_A \ot I_B + I_A \ot \log O_B$.

Similarly, we can show for the commutant that
\begin{align} \label{AlgRTCompAppendix}
S(\wt \rho, N') = S(\wt \rho, M') + \Tr (\wt \rho \cal L).
\end{align}
With $\cal {L}$ taken as the "area operator", we see that a Ryu-Takayani formula holds for any algebraic code with complementary recovery. Note that $S(\chi_\al, N) = S(\chi_\al, N')$ since $\chi_\al$ is a pure state, and that is why the area operators in \eqref{AlgRTAppendix} and \eqref{AlgRTCompAppendix} are the same.

We now want to prove the converse: that any theory obeying an exact algebraic Ryu-Takayanagi formula has complementary recovery. We first show that the "bulk" and "boundary" algebraic relative entropies are equal and then show that this implies condition (iii) of Theorem \ref{thrm:conditions}, which is equivalent to condition (ii).

Evaluating \eqref{AlgRTAppendix} for a small perturbation $\delta \wt \rho$ around a state $\wt \sigma$, we obtain
\begin{align} \nonumber
&S(\wt \sigma + \delta \wt \rho, N) = S(\wt \sigma + \delta \wt \rho, M) + \Tr ((\wt \sigma  + \delta \wt \rho)\cal L)\\ \nonumber
\implies &\Tr((\wt \sigma + \delta \wt \rho)K^{\wt \sigma}_N) = \Tr((\wt \sigma + \delta \wt \rho)K^{\wt \sigma}_M) + \Tr ((\wt \sigma  + \delta \wt \rho)\cal L) \\ \nonumber
\implies &\Tr(\delta \wt \rho K^{\wt \sigma}_N) = \Tr(\delta \wt \rho K^{\wt \sigma}_M) + \Tr (\delta \wt \rho \cal L)\\
\implies &\Tr(\wt \rho K^{\wt \sigma}_N) = \Tr(\wt \rho K^{\wt \sigma}_M) + \Tr (\wt \rho \cal L). \label{ModularHamiltonian}
\end{align}
Going from the first line to the second, we have used the definition of algebraic relative entropy given in eq. \eqref{RelativeEntropy} for the two states $\wt \sigma$ and $\wt \sigma + \delta \wt \rho$ and the fact that $S(\wt \sigma + \delta \wt \rho| \wt \sigma, N) = S(\wt \sigma + \delta \wt \rho| \wt \sigma, M) = 0$ to linear order in $\delta \wt \rho$. The third line follows from \eqref{AlgRTAppendix} and the forth line from integrating both sides of the equation. We now use \eqref{ModularHamiltonian} to show that for any two states $\wt \rho, \wt \sigma$ in the code subspace, we have
\begin{align}\nonumber
S(\wt \rho|\wt \sigma, N) &= -S(\wt \rho, N) + \Tr(\wt \rho K^{\wt \sigma}_N) \\ \nonumber
&= -S(\wt \rho, M) -\Tr(\wt \rho {\cal L}) + \Tr(\wt \rho K^{\wt \sigma}_N) \\ \nonumber
&= -S(\wt \rho, M) + \Tr(\wt \rho K^{\wt \sigma}_M)\\
&= S(\wt \rho|\wt \sigma, M),
\end{align}
and similarly for the commutant algebras.

Now suppose we take a state $\wt \rho$ and perturb it by  conjugating it by $e^{i \lambda \wt O}$, where $\wt O \in M$ is a Hermitian operator. In other words,
\begin{align}
    \wt \rho (\lambda) = e^{i \lambda \wt{O}} \wt \rho e^{-i \lambda \wt{O}}
\end{align}
Since $e^{i \lambda \wt O}$ commutes with any operator on $M'$, $\wt \rho_{M'} = \wt \rho_{M'}(\lambda)$. Hence
\begin{align}
    S(\wt \rho|\wt \rho(\lambda), N') = S(\wt \rho|\wt \rho(\lambda), M') = 0
\end{align}
and so the states $\wt \rho$ and $\wt \rho(\lambda)$ are indistinguishable on $N'$. It follows that for any operator $X' \in N'$, the variation of $\Tr (\wt \rho(\lambda) X')$ at linear order in $\lambda$ vanishes
\begin{align}
\Tr(\wt \rho [X',\wt O]) = 0.  
\end{align}
Since this is true for any operator $\wt O \in M$ and state $\wt \rho$ in the code space, we find that the operator $P_{code} X' P_{code}$, which has support only within the code space, commutes with any operator $\wt O \in M$. It can therefore be written as $\wt X' P_{code}$ for some operator $X' \in M'$, which is condition (iii) of Theorem \ref{thrm:conditions}.
\end{widetext}
\end{document}